# Molecular Doping of $MoS_2$ with N-Heterocyclic Carbenes: A Raman Study of Interfacial Charge Transfer

Ana Senkić[1,2,◊] Arne Nalop[3,◊], Xenia Sieverding[1,2,◊] , Ariane Ufer[1,2], Sebastian Roling[1,2], Frank Glorius[3,*], Ursula Wurstbauer[1,2,*]

[1] Institute of Physics, University of Münster, 48149 Münster, Germany
[2] Center for Soft Nanoscience (SoN), University of Münster, 48149 Münster, Germany
[3] Organisch-Chemisches Institut, University of Münster, 48149 Münster, Germany

◊ equally contributed
* corresponding: glorius@uni-muenster.de; wurstbauer@uni-muenster.de;

## ABSTRACT

We demonstrate efficient electron doping of exfoliated $MoS_2$ through the deposition of N-heterocyclic carbene (NHC) molecules onto its surface. Raman spectroscopy reveals a pronounced softening of the optical $A_{1g}$ phonon mode after NHC deposition, consistent with electron transfer from the NHC molecules to the $MoS_2$ layers. The largest phonon-energy shifts are observed in monolayer and bilayer flakes whereas changes in bulk-like $MoS_2$ are negligible, indicating that the doping efficiency depends on the number of layers. This thickness dependence highlights the dominant role of interfacial charge transfer at the NHC/$MoS_2$ interface. Quantitative analysis of the Raman modes of as-exfoliated and NHC-covered $MoS_2$, together with comparison with literature values, allows us to estimate an increase in the two-dimensional electron density of about two-orders of magnitude in $MoS_2$ mono- and bilayers. These results demonstrate that NHC-based surface deposition is a promising approach for chemically tuning the carrier density of semiconducting transition-metal dichalcogenides and provide a potential route toward molecularly tunable two-dimensional electronic devices, in which, analogous to modulation doping, the donors reside outside the semiconductor lattice.

## INTRODUCTION

The field of two-dimensional (2D) materials has undergone rapid development, driven by their distinctive physical properties and broad potential for technological applications. Among these materials, semiconducting transition metal dichalcogenides (TMDs), such as $MoS_2$, have emerged as particularly promising systems. In the monolayer limit, TMDs exhibit a direct bandgap [1,2], large exciton binding energies [3,4], and strong light-matter interactions [5,6], making them attractive for applications in photovoltaics, sensing, and optoelectronics at room temperature [7–10].

A key feature of $MoS_2$ is the high degree of tunability of its electronic and optical properties. This can be achieved through a variety of approaches, including strain engineering [11,12], electrostatic or chemical doping [13–15], and the assembly of mono- and few-layer structures into vertical hetero- and hybrid structures [16,17]. In addition, molecular deposition has recently gained attention as a versatile strategy for modifying the properties of $MoS_2$ and other semiconducting TMDs in a controlled manner [18–20]. Depending on the molecular structure and the interfacial electronic alignment, molecular deposition can enable several mechanisms for tailoring material behavior. These include defect passivation or modification of defect-related states [20–22], charge doping mediated by interfacial charge transfer [15,23,24], and, under resonant conditions, non-radiative energy transfer processes such as Förster resonance energy transfer (FRET) [20,25].

Recently, N-heterocyclic carbenes (NHCs) have emerged as a particularly suitable class of molecules for the controlled electron doping of TMDs such as $MoS_2$ [26,27]. NHCs offer strong $\sigma$-donor character, molecular tunability, and the ability to form self-assembled monolayers on gold [28–32] and other coinage metals [33–37], base metals [38–42], and selected semiconductors [43–46]. On such surfaces, covalent binding models [35,47–50], self-assembly [47,51,52], stability [28,53–55], competing adsorption geometries [31,37,56–58] and interfacial electron transfer [45,59–61] have been well explored. In contrast, the binding mode of NHCs on semiconducting TMDs is more complex. Proposed interaction mechanisms range from covalent binding between the carbene and the chalcogenide atoms [27,62], to frustrated Lewis acid-base pairing between the carbene and the metal atoms at defect sites [26], to physisorption of so-called masked carbenes [63].

Recent studies [27,28,63,64] indicate that NHC-derived species produce substantial interfacial electron doping in TMDs; however, the dominant binding mode and the charge transfer mechanism may depend on the molecular precursor, preparation conditions, defect density, and surface structure. Consequently, molecular deposition can affect the optical, electronic and vibrational properties, in particular the optical phonon modes through various contributions, including charge transfer, strain, and defect- or bonding-related perturbations. A quantitative comparison of as-exfoliated and NHC-covered $MoS_2$ samples is therefore required to distinguish these effects.

From an application perspective, controlled n-type doping of two-dimensional semiconductors is essential for reducing the contact resistance, adjusting the threshold voltage of field-effect transistors, and realizing complementary 2D logic and efficient photodetectors. Substitutional doping, however, introduces impurities into the channel lattice and thereby degrades the transport properties, whereas electrostatic gating requires additional device layers. In conventional semiconductors, modulation doping, in which the donors are spatially separated from the conducting channel, resolved this conflict and enabled high-mobility two-dimensional electron systems [65]. Molecular surface doping is a promising analogue for 2D semiconductors, because the donor molecules reside outside the lattice of the atomically thin channel and leave it unperturbed.

The central question addressed in this study is whether NHC deposition produces a reproducible electron-doping response in exfoliated $MoS_2$ that can be distinguished from strain, defect-related perturbations, and changes in the surrounding atmosphere.

In this work, we employ micro-Raman spectroscopy to investigate the impact of NHC deposition on exfoliated $MoS_2$ samples of varying thickness, ranging from monolayers to bulk-like flakes. From hyperspectral Raman maps acquired on three several micrometer sized samples containing mono-, bi-, few-layer and bulk-like regions, we perform a comparative statistical analysis of the optical phonon modes before and after NHC deposition. By quantitatively evaluating the mode frequencies and their relative shifts, we disentangle the contributions of changes in carrier density, strain, and defect- or covalent-bonding-related perturbations, building on previously established Raman-based analyses [5]. Our analysis reveals a pronounced n-type doping response after NHC deposition, which is strongest in monolayer and bilayer $MoS_2$ and negligible in bulk-like samples. The absence of Raman signatures of strain modification or defect generation after NHC deposition is consistent with a weak chemical perturbation of the $MoS_2$ lattice, although the binding mode cannot be determined from the present data. Comparison of the evolution of the $A_{1g}$ Raman-mode frequency in as-exfoliated and NHC-covered $MoS_2$ bilayers with literature values used for calibration [5,64,66] indicates an increase in the two-dimensional electron density by about two-orders of magnitude (up to $10^{13}$ $cm^{-2}$) due to interfacial charge transfer. This behavior resembles a molecular analogue of modulation doping [65], in which donor species located outside the $MoS_2$ lattice supply carriers to the conducting channel while leaving the lattice of the channel unperturbed. NHC deposition may therefore provide a route toward molecular doping in semiconducting TMDs for fundamental studies and optoelectronic applications.

# EXPERIMENTAL SECTION

**Sample preparation**

$MoS_2$ samples were cleaved by micromechanical exfoliation [67], dry-transferred onto $Si/SiO_2$ (300 nm) substrates, and characterized by micro-Raman measurements at room temperature using a commercial Raman microscope (Witec Alpha 300).

Prior NHC deposition, the $MoS_2$ samples were annealed in a vacuum chamber ($10^{-5}$ mbar) at 190°C for approximately one hour. The vacuum chamber and the physical vapor deposition (PVD) chamber were located in a glove bag in nitrogen atmosphere to ensure a controlled environment. After annealing, the $MoS_2$ samples were not exposed to air and were transferred to the PVD chamber within 5 minutes (the setup is described in the Supporting Information, Figure S11). The NHC precursor IMe·$CO_2$ was heated to 120 °C, and the vapor deposition time was about 30 minutes.

Along with the $MoS_2$ sample, a thin-film gold substrate was treated simultaneously as a control sample (Au-NHC). The Au-NHC sample was characterized via XPS (Figure S12 and Table S1). The successful deposition of NHCs on the gold surface proves the presence of free NHCs in the gas phase and, hence, the exposure of the$MoS_2$ samples to free NHC.

The NHC-covered $MoS_2$ samples were placed inside a small sample chamber (Linkam Cell) filled with dry nitrogen gas directly evaporated from the liquid phase for highest purity. In this way, the Raman measurements of the NHC-covered samples were done in inert nitrogen atmosphere.

**Optical measurements**

All optical measurements were conducted on a commercial Raman confocal microscope (Witec Alpha 300) with 488 nm (2.54 eV) laser excitation and a diffraction-limited spot size. For Raman measurements, a grating with 1800 lines per mm was used. For as-exfoliated $MoS_2$ (NHC covered $MoS_2$) samples, a Neofluar 100x (50x) objective with 1 (9.1) mm working distance was employed, with laser power up to 1 mW (2 mW) and integration time up to 2 s (1.5 s). The laser power was increased for the NHC-deposited $MoS_2$ samples to maintain a comparable power density, accounting for the different objective and optical losses introduced by the cell window. No degradation or laser-induced modifications were observed in any of the samples, consistent with the low power densities used. Since the analysis focused exclusively on the Raman peak frequencies and full widths at half maximum (FWHMs), rather than on peak intensities, variations in laser power do not affect the reported parameters. For the as-exfoliated $MoS_2$ sample, additional measurements conducted in ambient and $N_2$ atmospheres were performed with a Neofluar 50x objective with a 9.1 mm working distance, a laser power of 2 mW and integration time of 2 s. All measurements were conducted at room temperature. Because the as-exfoliated and NHC-covered samples were measured with different objectives and laser powers, the bulk-like region of each flake serves as an internal reference. For simplicity, the bulk notation $A_{1g}$ and $E_{2g}$ is used for all thicknesses.

# RESULTS AND DISCUSSION

A schematic illustration of NHC molecules deposited on the surface of few-layer $MoS_2$ sample is shown in Figure 1a. Figure 1b shows typical Raman spectra of exfoliated monolayer (1L), bilayer (2L), and few-layer (FL) $MoS_2$ samples before and after NHC deposition (sample shown in Figure 3). Here, the in-plane $E_{2g}$ phonon mode is centered at a Raman shift of (386.19 ± 0.03) $cm^{-1}$ and the out-of-plane $A_{1g}$ phonon mode at a Raman shift of (405.26 ± 0.05) $cm^{-1}$ for the as-exfoliated $MoS_2$ monolayer [68–70]. As depicted in Figure 1b and elaborated in detail later on, the frequency separation between $A_{1g}$ and $E_{2g}$ phonon modes was found to decrease after NHC deposition due to softening of the $A_{1g}$ mode, while the $E_{2g}$ mode shows no shift relative to the as-exfoliated sample. This $A_{1g}$ softening after NHC deposition depends on layer number.

To quantify this energy shift, all Raman spectra were fitted to a sum of two Lorentzian functions, from which information about their frequency shift, intensity, and full width at half maximum was extracted. With increasing number of layers, the frequency of the $A_{1g}$ mode increases due to stiffening of the lattice, while the frequency of the $E_{2g}$ mode decreases due to the combined effect of enhanced dielectric screening of the long-range Coulomb interaction and interlayer hybridization effect [71]. The opposite trend in their thickness-dependent frequency makes the frequency separation of these two modes a reliable indicator of the $MoS_2$ layer number [5,70,72]. Additional layer-number dependent Raman data are as shown in Figures S1 a,b). In the monolayer limit, the frequency difference between these two modes is below 20 $cm^{-1}$, whereas for bilayers it increases to 20-22 $cm^{-1}$; for bulk-like samples, the frequency difference is $\geq$ 25 $cm^{-1}$. These values provide useful boundaries, particularly for distinguishing monolayers and bilayers, although the exact values depend on the charge-carrier density [5,64,66] and other parameters such as strain and defects. It is worth mentioning that this method for determining the number of layers is valid for exfoliated samples, whereas for as-grown samples the frequency difference has been reported to depend on the synthesis parameters [73].

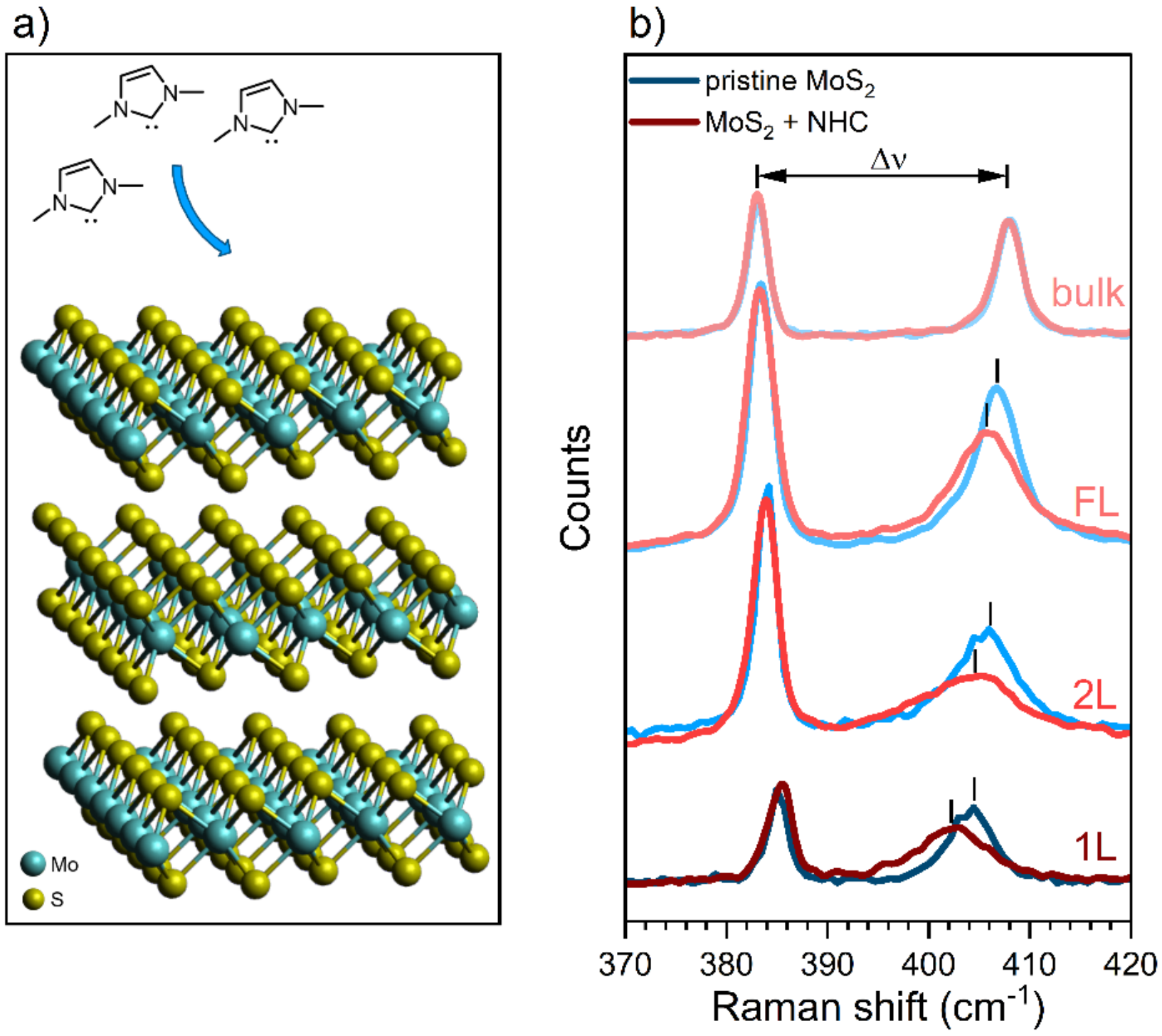

**Figure 1.** a) Schematic representation of NHC deposition on a few-layer $MoS_2$ sample. Image was generated based on crystallographic $MoS_2$ data (ID mp-2815) using Avogadro-1.2.0 and ChemDraw-25.0.2. b) Raman spectra acquired from mono- (1L), bi- (2L), few-layer (FL) and bulk regions of sample 2 before and after NHC deposition (positions marked by X in Figure 3). Spectra are normalized to the $E_{2g}$ phonon mode intensity and vertically offset for clarity. Frequency separation $\Delta\nu$ between $A_{1g}$ and $E_{2g}$ phonon modes decreases after NHC deposition due to softening of the $A_{1g}$. This relative shift depends on the layer-number. Softening of the $A_{1g}$ optical phonon mode together with increased FWHM are indicative of an increased electron-density due to charge transfer from the NHC molecules.

Beyond their thickness dependence, the frequencies and FWHM of these two phonon modes are also sensitive to external perturbations including strain [74,75], defects [66,76], and changes in the electron density induced by electronic or structural dopants [20,64,77–79], as illustrated in Figures S1 c) and S2. In particular, changes in electron density can modify the phonon frequencies of the $A_{1g}$ mode through electron–phonon coupling [66] (see also Figure S2). Figure 2 a) shows an optical micrograph of an exfoliated multilayer $MoS_2$ flake, in which regions with different layer thicknesses can be identified from their optical contrast arising from the interference effects [80]. Regions with 2L and FL thickness are outlined with dashed white lines. Most of the flake exhibits a bulk-like thickness and appears yellow. Figures 2 b) and c) show spatial maps of the frequency separation $\Delta\nu = \nu(A_{1g}) - \nu(E_{2g})$ between $A_{1g}$ and $E_{2g}$ phonon modes for the $MoS_2$ flake (sample 1) before and after NHC deposition, respectively. Through careful analysis of the $E_{2g}$ and $A_{1g}$ mode frequencies in the pristine and NHC-deposited flake, we identify a clear redshift for the $A_{1g}$ mode after NHC deposition, while the $E_{2g}$ mode shows no systematic shift. The absence of a measurable shift in the $E_{2g}$ mode together with the absence of the defect-activated LA(M) mode (~227 $cm^{-1}$ [66] ) largely rules out strain- and defect-induced effects and indicates that the observed changes predominantly originate from the electronic modulation. Accordingly, we attribute the observed reduction in $\Delta\nu$ to the enhanced electron doping resulting from charge transfer from the NHC molecules to $MoS_2$. This interpretation is consistent with the opposite Raman response reported for adsorbed water molecules, which deplete $MoS_2$ of electrons via molecular gating [64]. This interfacial charge-transfer mechanism is conceptually related to modulation doping, in which spatially separated donor species transfer carriers to a two-dimensional conducting channel, while the lattice of the channel remains unperturbed. The modulation-doping approach was originally demonstrated in GaAs/AlGaAs heterostructures, where it enabled the formation of high-mobility two-dimensional electron systems [65,81].

Following NHC deposition, the bulk-like region of the $MoS_2$ sample remains largely unchanged, whereas the thinner regions exhibit a reduction in optical phonon frequency separation $\Delta\nu$ of between 1-2 $cm^{-1}$, depending on the flake's thickness. This observation is consistent with the expected decrease in the molecular-doping effect with increasing layer number N, since the maximum amount of interfacial charge transfer per unit area $n$ is limited by the molecular coverage [64]. Consequently, the transferred electron-density per layer approximately follows the relation $n$/N, therefore reducing the doping-induced phonon-frequency shift and making it unresolvable in thicker flakes [64]. We emphasize that a decrease in the frequency separation $\Delta\nu$ is indicative for n-type doping [20,64,77,82]. This interpretation is further supported by the mode-resolved maps in Figure S3: only the $A_{1g}$ mode exhibits a pronounce softening [83–85].

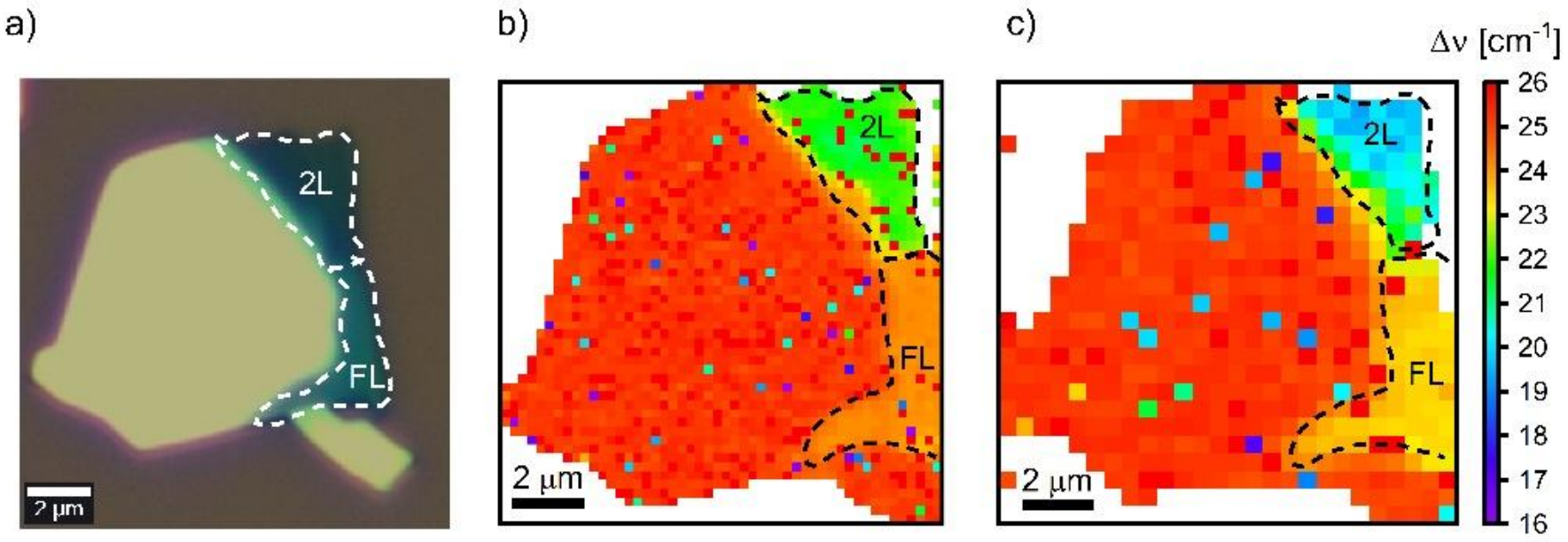

**Figure 2.** a) Optical micrograph of an exfoliated multilayer $MoS_2$ sample, with bilayer (2L) and few-layer (FL) regions marked with dashed line. b), c) Spatial maps of the frequency separation $\Delta\nu$, between the $E_{2g}$ and $A_{1g}$ modes of sample 1 before and after NHC deposition, respectively.

It is also worth noting that, in addition to the frequency softening, the FWHM of the $A_{1g}$ mode increases after NHC deposition, from approximately 7 $cm^{-1}$ to 15 $cm^{-1}$ in the bilayer region, while the FWHM of $E_{1g}$ mode shows no systematic changes. The broadening of $A_{1g}$ mode is consistent with substantial n-type doping, although inhomogeneous broadening may also contribute [64,77].

To verify the reproducibility of the effect of NHC deposition on the vibrational modes of $MoS_2$, we performed measurements on two additional exfoliated samples (samples 2 and 3) each containing regions of flakes with different thicknesses, as shown in Figures 3 and 4. These measurements confirm that NHC deposition results in shifts of the Raman frequencies with a thickness dependence similar to that observed in the initial analysis: monolayer $MoS_2$ regions show a larger change in the frequency separation $\Delta\nu$ between as-exfoliated and NHC-deposited flakes than thicker regions (Figures 3c and 4c). Based on the change in the frequency separation Δν, the resulting change in charge-carrier density can be estimated using an established calibration [64]. For mono- and bilayer $MoS_2$, changes in Δν of approximately 2 $cm^{-1}$ and 1.2-1.5 $cm^{-1}$, respectively, are observed. For monolayers, the calibration in Figure S1c yields an increase in electron density of $\Delta n \approx 9 \times 10^{12}$ $cm^{-2}$, or $\approx 7 \times 10^{12}$ $cm^{-2}$ after subtracting the atmosphere contribution (Figure 5). Such shifts are consistently observed in all investigated $MoS_2$ samples after NHC deposition. The frequency changes of the $E_{2g}$ and $A_{1g}$ modes before (as-exfoliated) and after NHC deposition for three samples are shown in Figures S3, S5, S7 in the SI. Finally, the changes in the FWHMs of both modes before and after NHC depositions are shown in Figures S4, S6 and S8, with a significant broadening of $A_{1g}$ mode after NHC deposition.

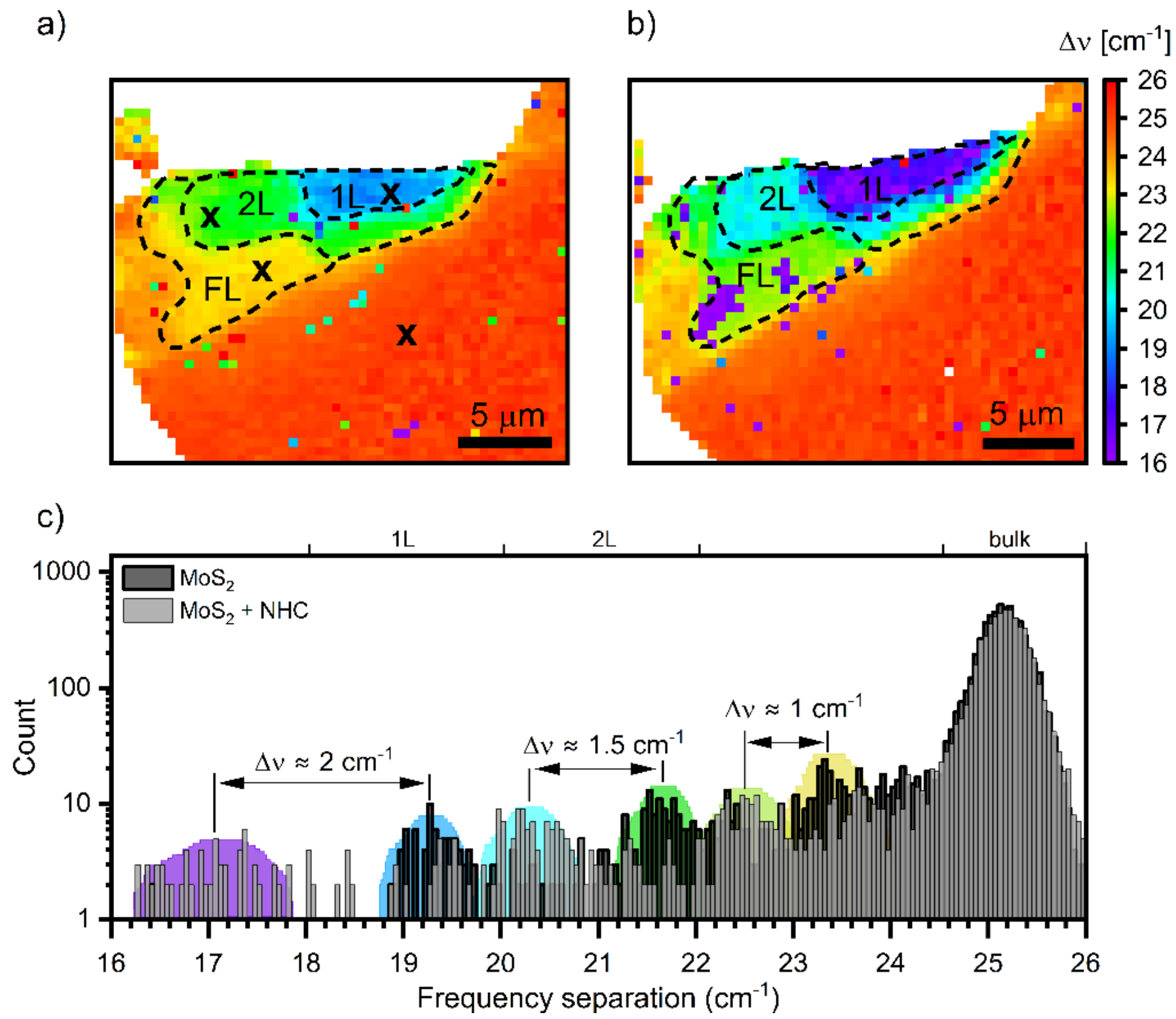

**Figure 3.** a), b) Spatial maps of the frequency separation Δν between the $E_{2g}$ and $A_{1g}$ modes for sample 2 before and after NHC deposition, respectively. Dashed lines outline regions with monolayer (1L), bilayer (2L), and few-layer (FL) thicknesses. The "X" symbols mark the positions at which the point spectra shown in Figure 1 were acquired. c) Histograms of frequency separation Δν between $E_{2g}$ and $A_{1g}$ modes before and after NHC deposition. The literature ranges for monolayer (1L), bilayer (2L), and bulk-like samples are indicated by vertical lines on the top axis [68–70].

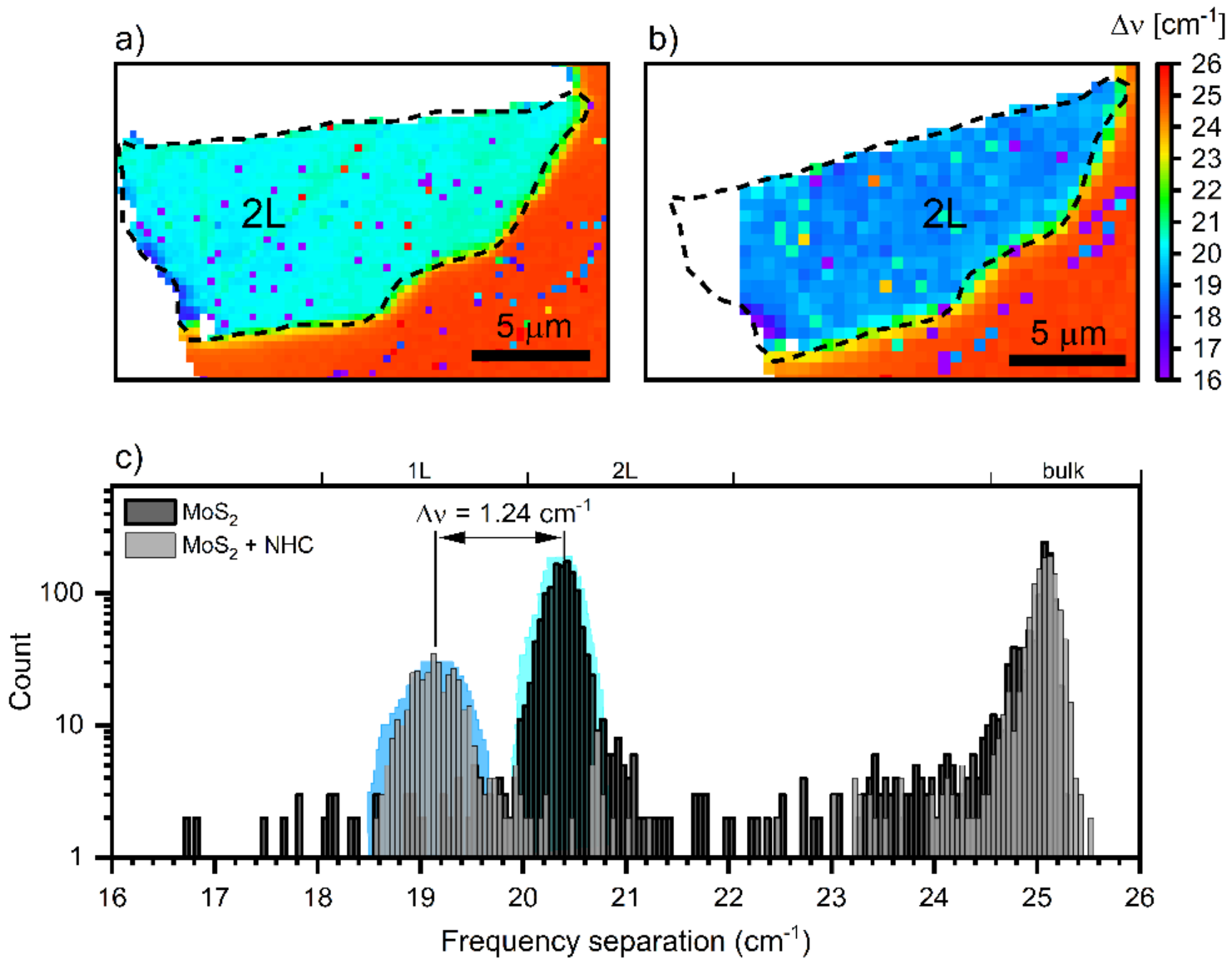

**Figure 4.** a), b) Spatial maps of the frequency separation Δν between the $E_{2g}$ and $A_{1g}$ modes for sample 3 before and after NHC deposition, respectively. Dashed lines outline regions with bilayer (2L) thickness. c) Histograms of frequency separation Δν between $E_{2g}$ and $A_{1g}$ modes before and after NHC deposition. The literature ranges for monolayer (1L), bilayer (2L), and bulk-like samples are indicated by vertical lines on the top axis [68–70].

To assess whether the surrounding atmosphere influences the intralayer vibrational modes of $MoS_2$ and to distinguish environmental effects from NHC-induced doping, we performed a complementary Raman experiment under ambient and nitrogen atmospheres. In this comparison, a separate as-exfoliated $MoS_2$ reference sample (not NHC-treated) was measured first under ambient conditions, which include exposure to water vapor and oxygen, and subsequently under an inert $N_2$ gas atmosphere [64] using the same measurement configuration as for the NHC-covered flakes. As shown in Figure 5, switching from ambient air to $N_2$-gas decreases the frequency separation Δν by only 0.5 $cm^{-1}$ in the monolayer and bilayer region, while the frequency separation of the bulk-like region remains unchanged. The reduced frequency separation observed under an $N_2$ atmosphere is attributed to a slight increase in the charge-carrier density resulting from reduced depletion, which is predominantly caused by molecular gating by water molecules present in the ambient atmosphere [64]. These comparatively small changes are substantially less pronounced than those observed after NHC deposition. Since the NHC-covered samples were measured in $N_2$ and the as-exfoliated samples in ambient air, this contribution of ≈ 0.5 $cm^{-1}$ is contained in the shifts discussed above; the NHC-induced change thus amounts to ≈ 1.5 $cm^{-1}$ for monolayers. The control experiment therefore indicates that changing the atmosphere from ambient air to $N_2$ has only a minor effect on the Raman response and that the dominant contribution to the $A_{1g}$ softening, indicative for electron doping, arises from interfacial charge transfer between the NHC molecules and $MoS_2$. Thus, NHC deposition provides an effective route for electron doping of $MoS_2$ through molecularly induced interfacial charge transfer.

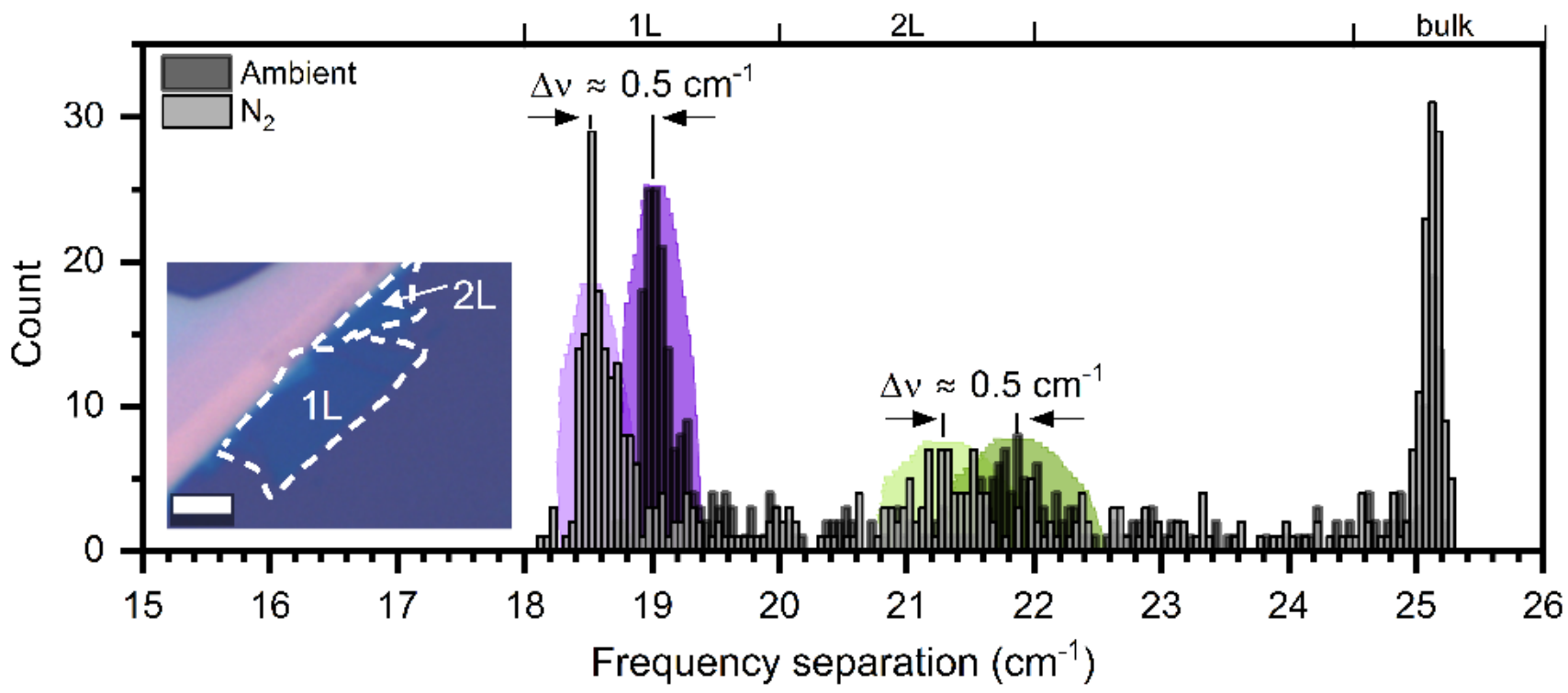

**Figure 5.** Histograms of frequency separation between $E_{2g}$ and $A_{1g}$ modes of an as-exfoliated $MoS_2$ reference sample in $N_2$ and ambient atmosphere. The literature ranges for monolayer (1L), bilayer (2L), and bulk-like samples are indicated by vertical lines on the top axis [68–70]. Inset in the bottom left corner shows an optical image of the exfoliated $MoS_2$ sample. Dashed lines outline regions with monolayer (1L), and bilayer (2L) thicknesses. Scalebar is 4μm.

## CONCLUSIONS

In summary, a comparative Raman analysis of exfoliated $MoS_2$ samples before and after NHC deposition with $MoS_2$ thicknesses ranging from monolayers to bulk-like flakes reveals pronounced n-type doping after NHC deposition. The dominant response is a red-shift and broadening of the $A_{1g}$ mode, whereas the $E_{2g}$ mode shows no systematic shift before and after NHC deposition. Together with the absence of substantial changes attributable to strain or defect generation, these observations are consistent with electron transfer from the deposited NHC molecules to $MoS_2$ and with limited chemical perturbation of the lattice. The effect is strongly thickness dependent: frequency shifts of up to approximately 2 $cm^{-1}$ are observed in monolayer regions, whereas the $A_{1g}$–$E_{2g}$ frequency separation in bulk-like regions remains essentially unchanged. For monolayer $MoS_2$, the change in the $A_{1g}$–$E_{2g}$ frequency separation corresponds, based on established Raman calibrations, to an increase in the electron density on the order of $10^{13}$ $cm^{-2}$ (typically two-orders of magnitude). This trend is reproduced across multiple samples. Measurements of an as-exfoliated $MoS_2$ reference sample under ambient and $N_2$ atmospheres show substantially smaller changes, indicating that the dominant contribution to the observed doping arises from NHC-induced interfacial charge transfer rather than from the change in the surrounding atmosphere. The NHC/$MoS_2$ system therefore provides a molecular charge-transfer-doping scheme that is conceptually related to modulation doping, although the donor layer is directly adjacent to the $MoS_2$ rather than separated by a conventional spacer layer. Because the donors reside outside the $MoS_2$ lattice, NHC deposition offers a promising route for molecular control of the carrier density in semiconducting TMDs that leaves the channel lattice unperturbed, which is relevant for contact engineering, field-effect transistors, and optoelectronic devices.

## ASSOCIATED CONTENT

Supporting Information: Layer-dependent Raman reference data and Raman spectra of 1L $MoS_2$ in different environments; Raman frequency and FWHM maps of samples 1–3; synthesis and NMR characterization of IMe·$CO_2$; physical vapor deposition setup; XPS of IMe on Au (PDF)

## ACKNOWLEDGEMENTS

The authors gratefully acknowledge financial support by the German Science Foundation (DFG) via Grants 443274199 and 556436549 (WU 637/7-2,8-1), No.426726249 (DE 2749/2-1 and DE 2749/2-2), and the Priority Program 2244 (2DMP). A.S. acknowledges the Alexander von Humboldt Foundation for a research fellowship. During the preparation of this work, the authors used large language models to refine the manuscript regarding grammar, readability, and clarity; the authors reviewed and edited the content and take full responsibility for the publication.

## DATA AVAILABILITY STATEMENT

The data that support the findings of this study are available from the corresponding author upon reasonable request.

*Supporting Information*

# Molecular Doping of $MoS_2$ with N-Heterocyclic Carbenes: A Raman Study of Interfacial Charge Transfer

Ana Senkić[1,2,◊] Arne Nalop[3,◊], Xenia Sieverding[1,2,◊] , Ariane Ufer[1,2], Sebastian Roling[1,2], Frank Glorius[3,*], Ursula Wurstbauer[1,2,*]

[1] Institute of Physics, University of Münster, 48149 Münster, Germany
[2] Center for Soft Nanoscience (SoN), University of Münster, 48149 Münster, Germany
[3] Organisch-Chemisches Institut, University of Münster, 48149 Münster, Germany

◊ equally contributed
* corresponding: glorius@uni-muenster.de; wurstbauer@uni-muenster.de;

## Additional layer-dependent Raman spectra

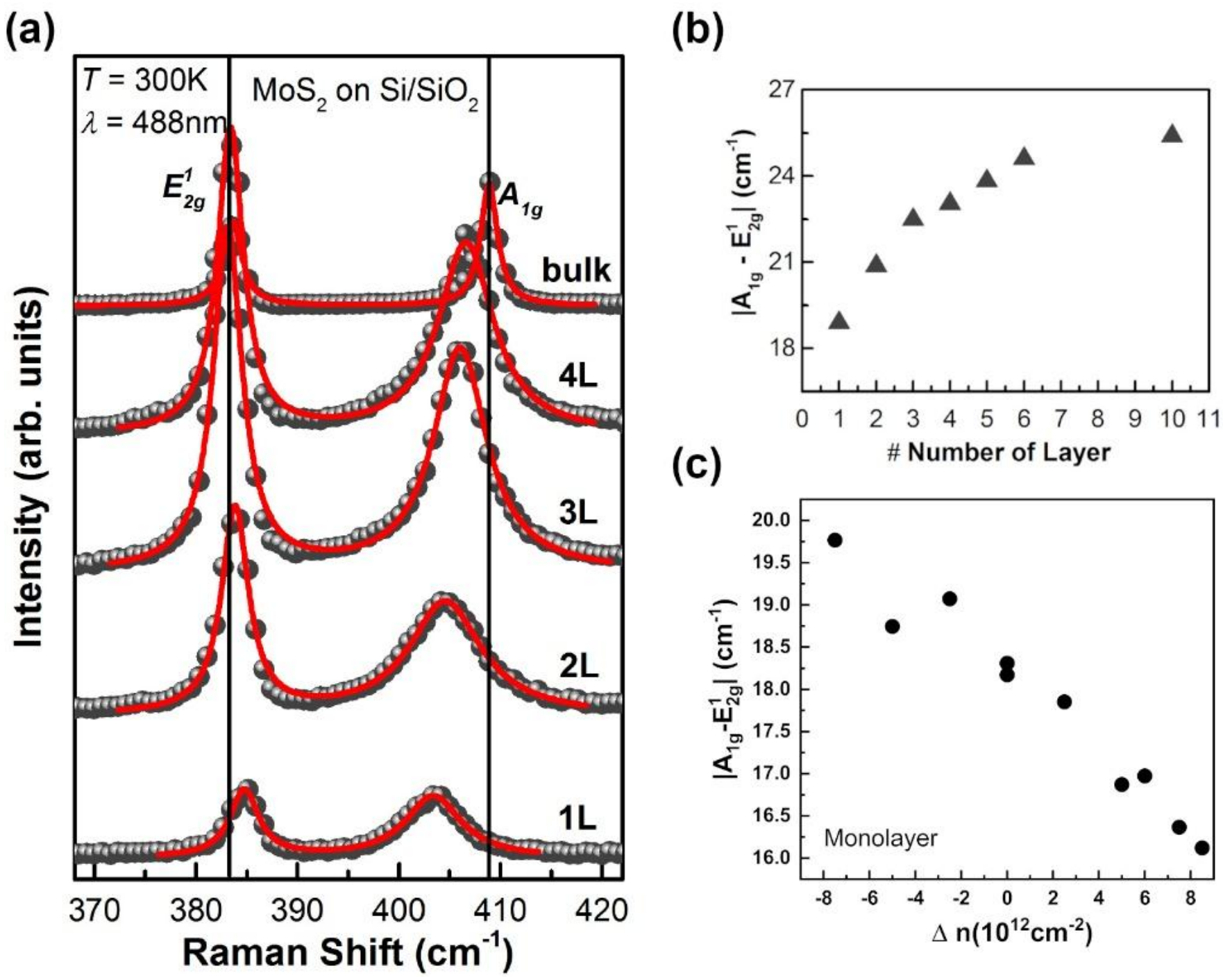

**Figure S1.** a) Layer-dependent Raman spectra of an additional as-exfoliated $MoS_2$ reference sample taken at room temperature with 488nm (2.54 eV) excitation. b) Frequency separation of $A_{1g}$ and $E_{2g}$ modes as a function of number of layers in $MoS_2$. c) Frequency separation of the $A_{1g}$ and $E_{2g}$ modes as a function of the change in charge-carrier density in monolayer $MoS_2$ measured in dependence of a solid electrolyte gate voltage (similar as described in [65]).

## Raman spectra on 1L $MoS_2$ in different environments: impact of adsorbed environmental molecules

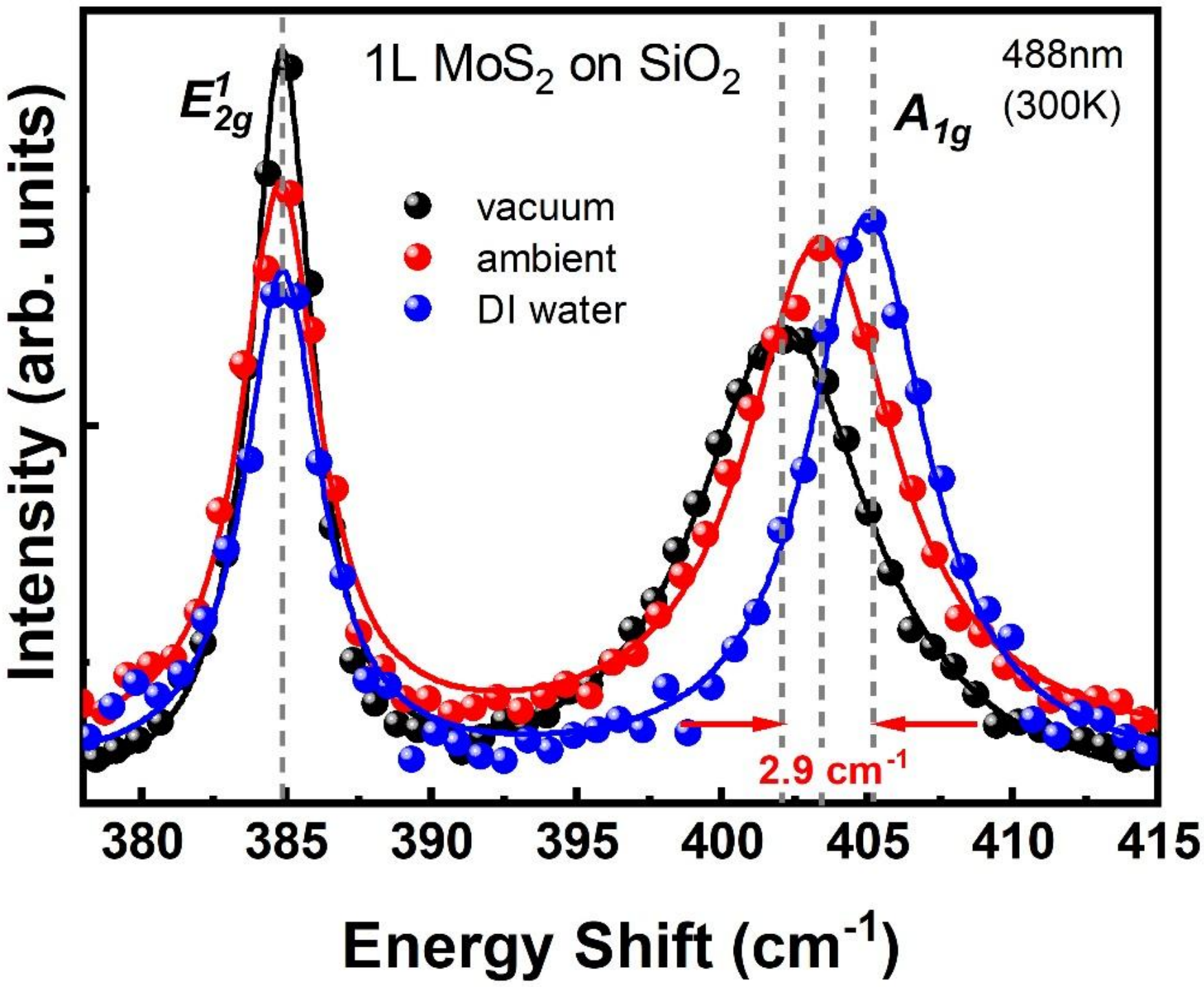

**Figure S2.** a) Raman spectra of the $E_{2g}$ and $A_{1g}$ modes under different conditions (vacuum, ambient and deionized water) in monolayer (1L) $MoS_2$ at room temperature.

## Additional Raman maps of sample 1 (Figure 2) before and after NHC deposition

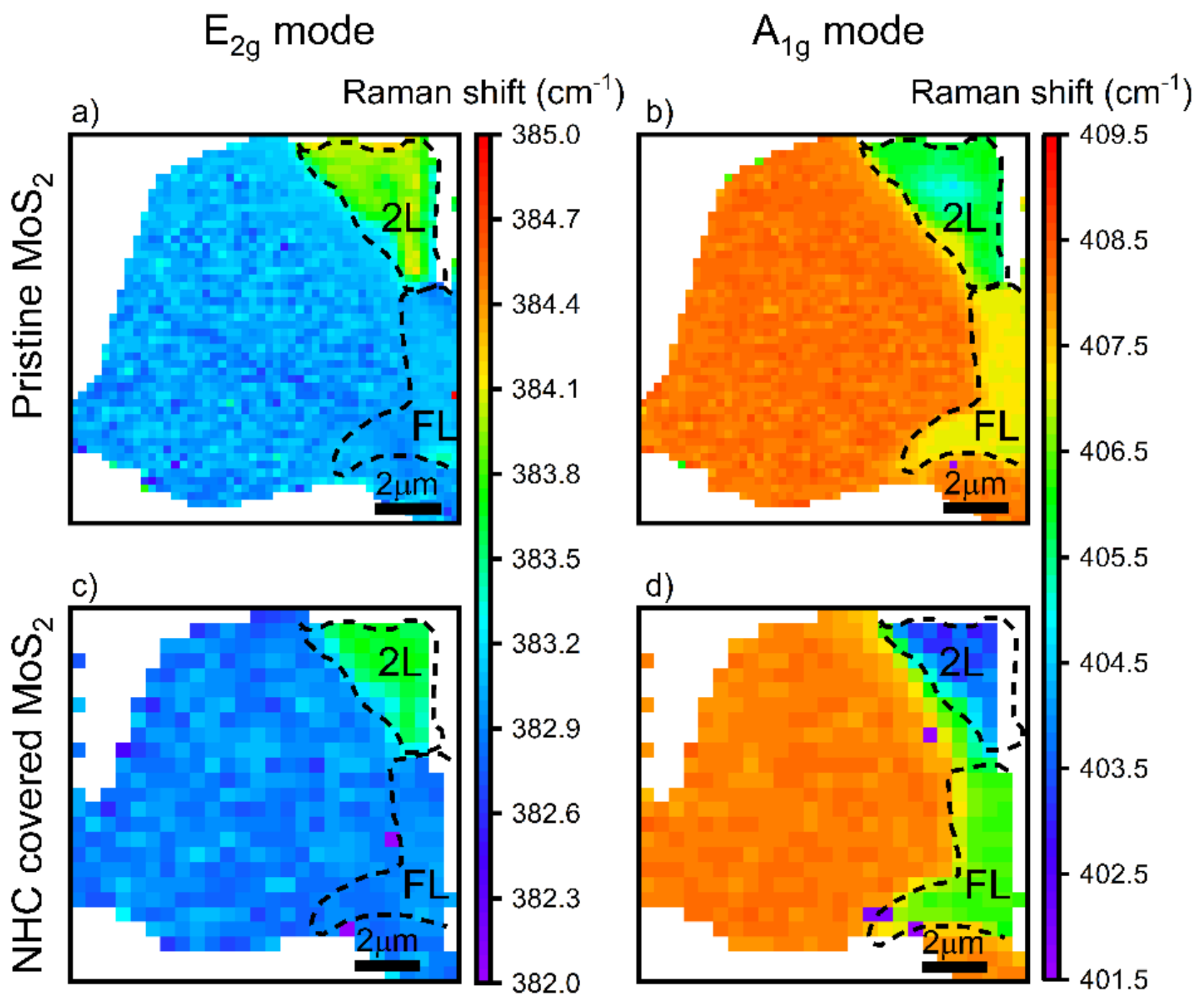

**Figure S3.** Frequencies of $E_{2g}$ and $A_{1g}$ modes of sample 1 before (a,b) and after (c,d) NHC deposition.

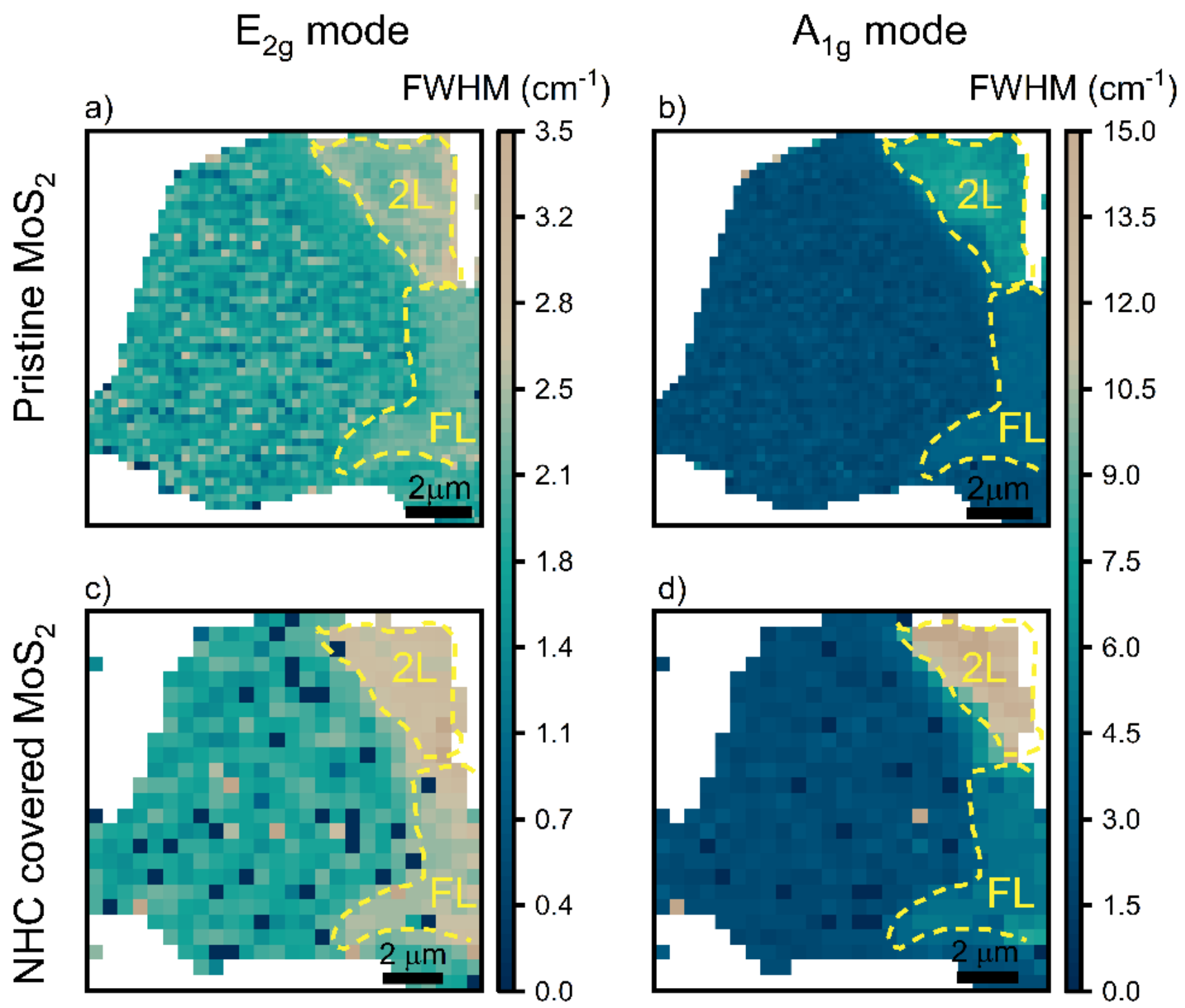

**Figure S4.** FWHM of $E_{2g}$ and $A_{1g}$ modes of sample 1 before (a,b) and after (c,d) NHC deposition.

## Additional Raman maps of sample 2 (Figures 1 and 3) before and after NHC deposition

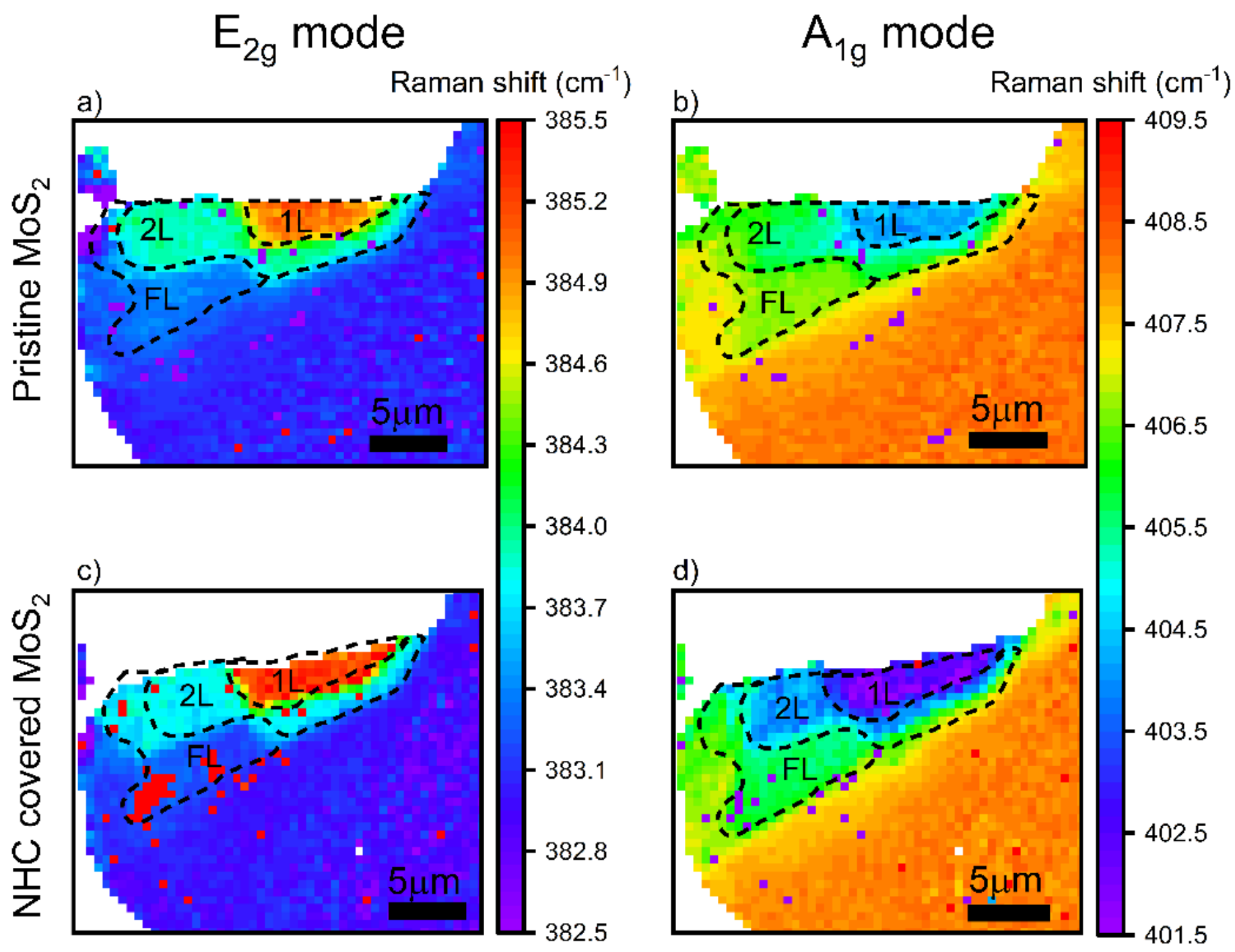

**Figure S5.** Frequencies of $E_{2g}$ and $A_{1g}$ modes of sample 2 before (a,b) and after (c,d) NHC deposition.

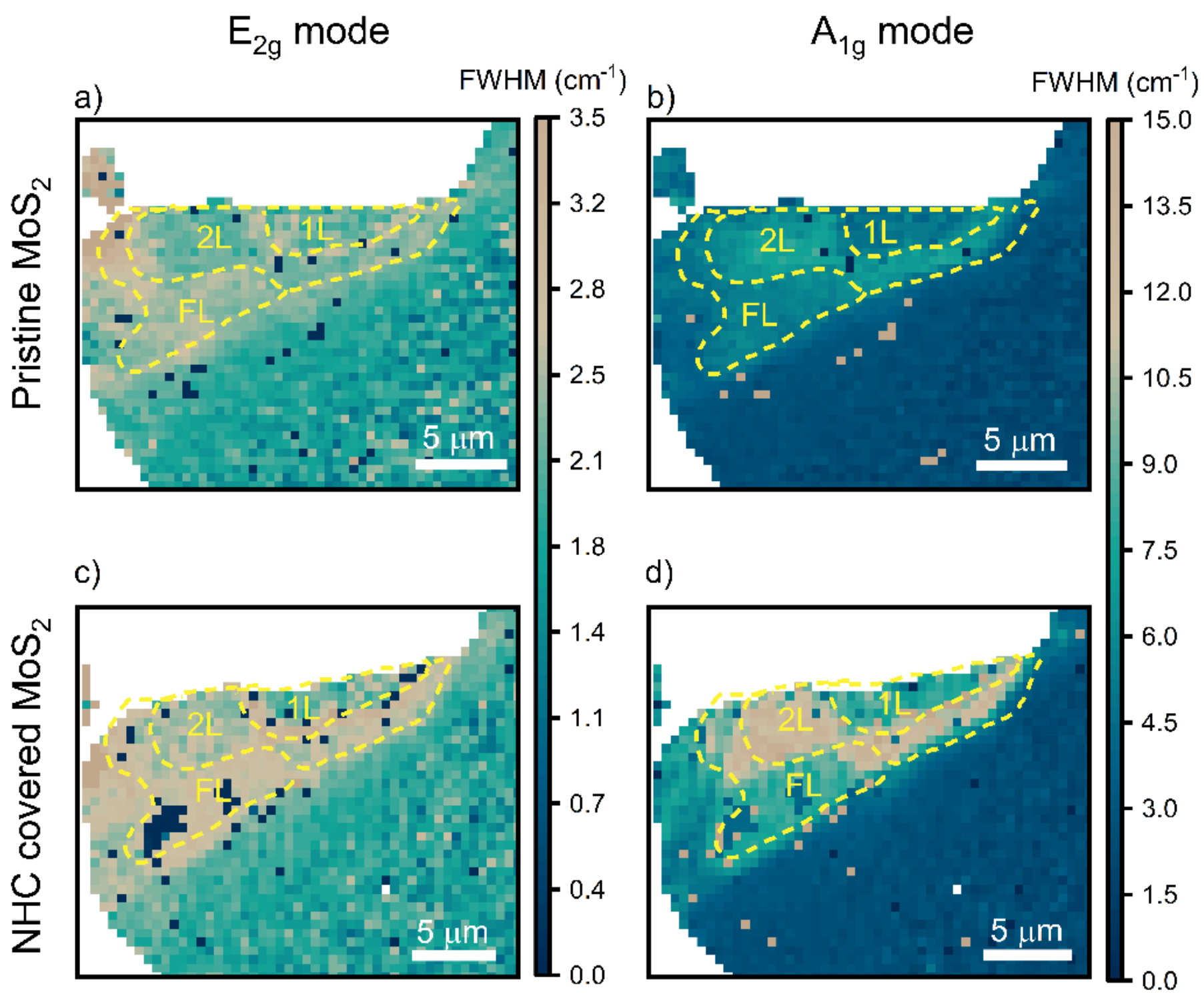

**Figure S6.** FWHM of $E_{2g}$ and $A_{1g}$ modes of sample 2 before (a,b) and after (c,d) NHC deposition.

## Additional Raman maps of sample 3 (Figure 4) before and after NHC deposition

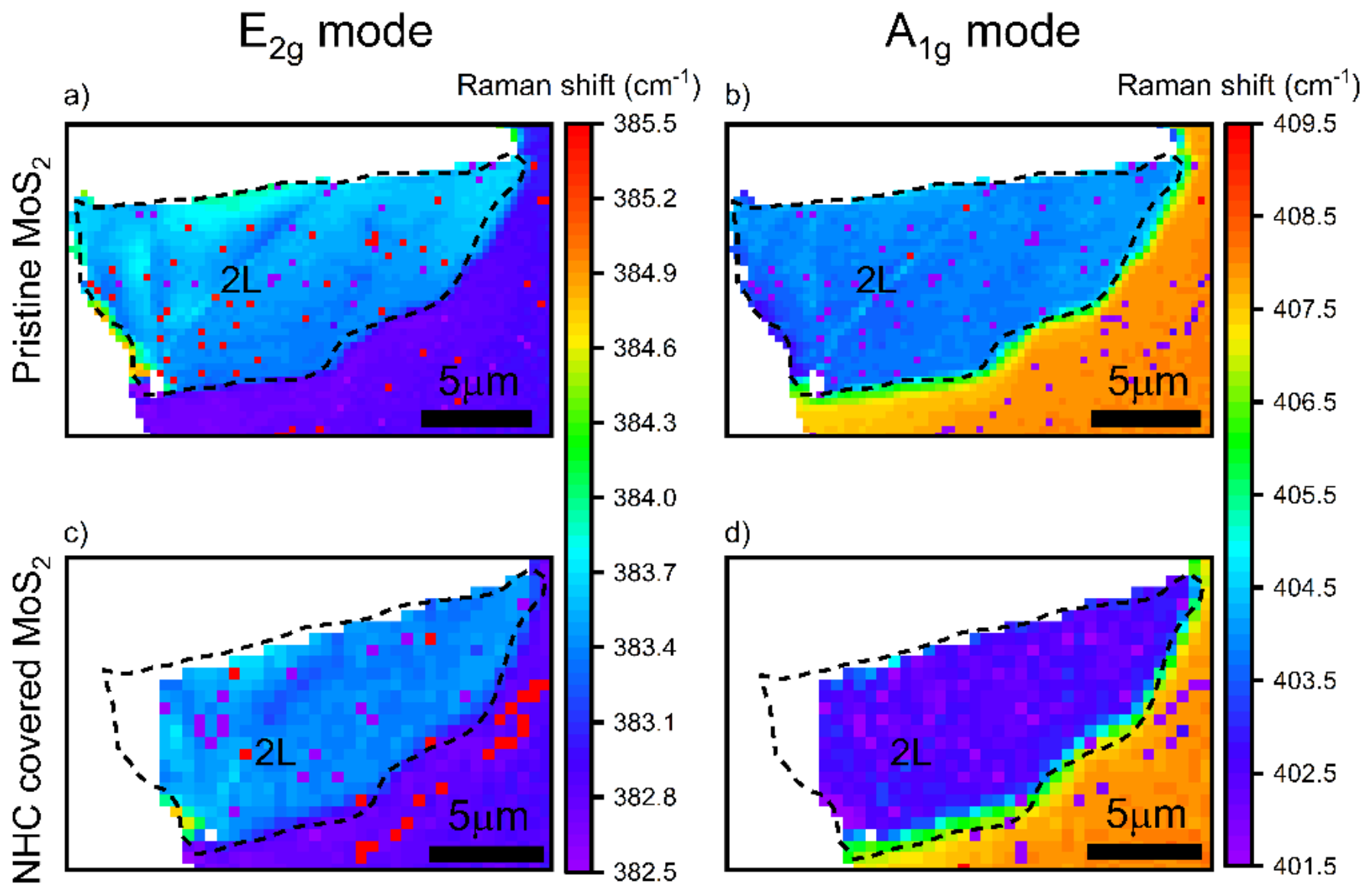

**Figure S7.** Frequencies of $E_{2g}$ and $A_{1g}$ modes of sample 3 before (a,b) and after (c,d) NHC deposition.

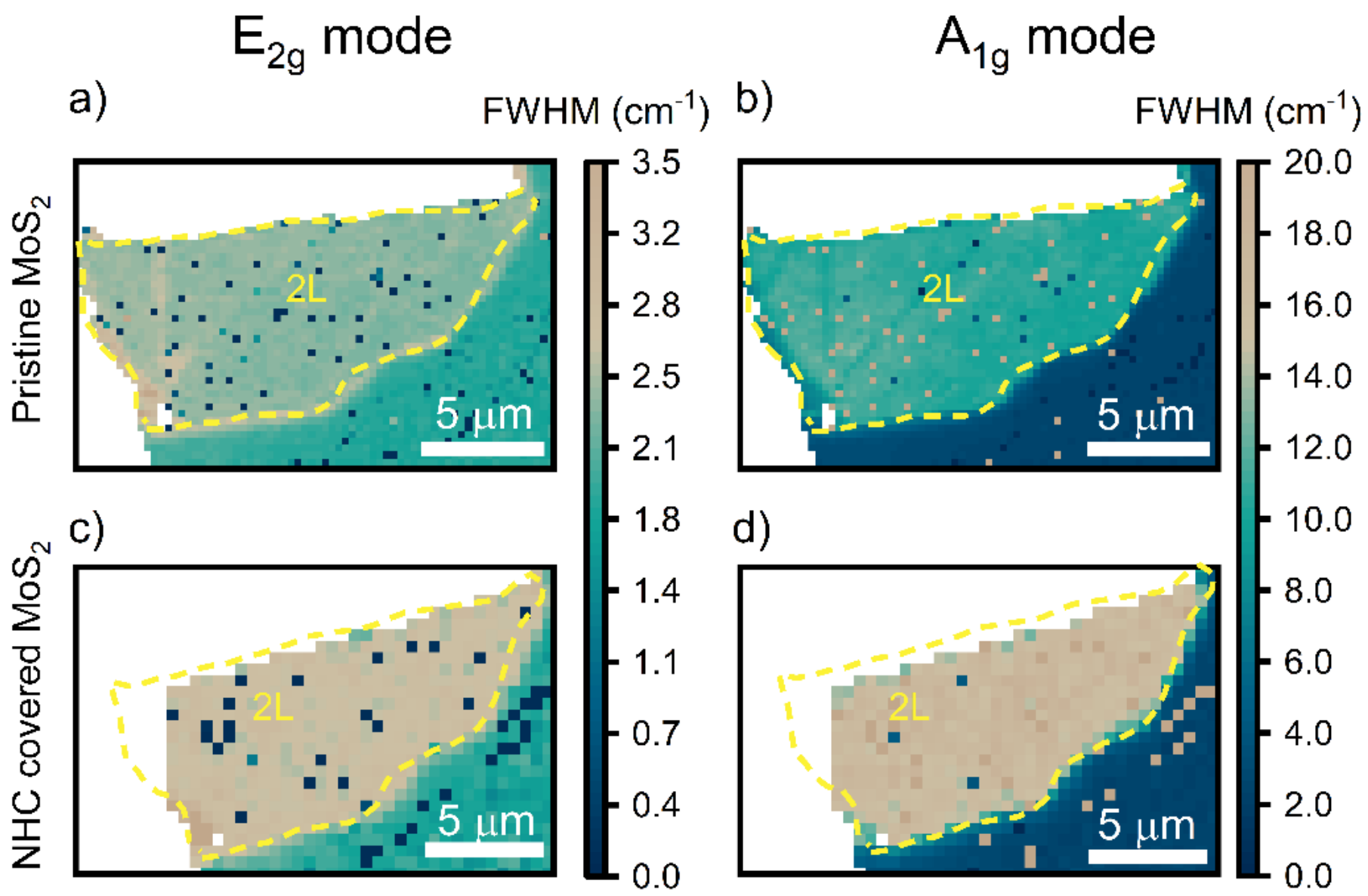

**Figure S8.** FWHM of $E_{2g}$ and $A_{1g}$ modes of sample 3 before (a,b) and after (c,d) NHC deposition.

## Chemical Synthesis

Reactions were stirred using oven-dried PTFE-coated magnetic stir bars with a rotation speed of 750 rpm. Solids were weighted under ambient conditions except otherwise mentioned, liquids handled with pipettes or syringes. Reaction temperatures are reported as the temperature of metal heating blocks or cooling baths surrounding the reaction vessel. The following solvents were purchased from ACROS Organics, Sigma-Aldrich and Fischer Scientific and used from a solvent purification system, which dried HPLC grade solvents over alumina under a positive pressure of argon according to Grubb's procedure: Tetrahydrofuran (THF), toluene and dichloromethane ($CH_2Cl_2$). Other solvents were also purchased from ACROS Organics, Sigma-Aldrich and Fischer Scientific and stored over activated molecular sieves. Solvents for standard work-up or purification procedures were simply purified by distillation (*n*-pentane, $CH_2Cl_2$, EtOAc, $Et_2O$).

Starting materials were bought from commercial suppliers such as ABCR, Acros Organics, Alfa Aesar, BLD Pharm, Carbolution, Combi-Blocks, Fisher Scientific, Fluka, Fluorchem, Merck, Sigma-Aldrich, Strem Chemicals, or TCI and used as received. Air or moisture sensitive compounds were stored and handled in an argon filled glove-box with oxygen and water levels below 1 ppm.

$^{1}$H- and $^{13}$C-NMR spectra were recorded at room temperature on an Avance II 400 MHz instrument. Chemical shifts (δ) are given in ppm. Residual solvent signals were used as references for 1H-NMR spectra: $D_2O$ ($\delta_H$ 4.79 ppm). Spectra were analyzed with MestReNova from Mestrelab Research S.L. and multiplicity described with standard abbreviations.

ESI mass spectra (HRMS, high resolution mass spectrometry) were recorded on a Thermo Fisher Scientific Orbitrap LTQ XL or a Thermo Fisher Scientific Orbitrap Velos Pro.

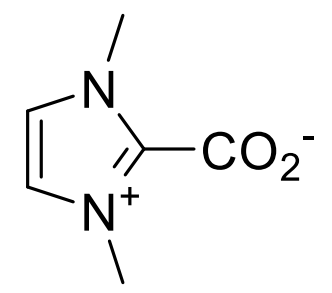

1,3-dimethyl-1*H*-imidazol-3-ium-2-carboxylate

1,3-dimethyl-1H-imidazolium iodide (IMeH·I) was synthesized according to a known literature procedure.[1,2] 1,3-dimethyl-1H-imidazol-3-ium-2-carboxylate (IMe·$CO_2$) was synthesized according to adapted literature procedures:[1,2] IMeH·I (10 g, 44.6 mmol, 1 eq.) and potassium *tert*-butoxide (6.0 g, 53.6 mmol, 1.2 eq., weighed under inert argon atmosphere) were placed in a oven dired Schlenk flask under argon. At -78 °C, 250 ml dry and degassed tetrahydrofuran (THF) were added. Whilst warming slowly to rt., the mixture was stirred for 2.5 hours. The mixture was filtered over Celite under argon. The filter was washed with THF (two times 250 ml). Still under argon, $CO_2$ was infiltrated into the yellow carbene solution and white precipitate was formed. After one hour, the white precipitate was filtered under ambient condition, washed with 100 ml THF and 400 ml $CH_2Cl_2$, sonicated in MeOH, washed again with $CH_2Cl_2$ and *n*-pentane, dried under reduced pressure and stored inside a glovebox fridge (4.8 g, 34.0 mmol, 76 % yield).

$^{1}$H NMR (400 MHz, $D_2O$) δ = 7.36 (s, 2H), 3.98 (s, 6H).

$^{13}$C NMR (101 MHz, $D_2O$) δ = 160.53, 158.36, 123.05, 36.73.

Pushrod (thermal evaporation coupled to EI): m/z calculated for $[IMe]^{+\cdot}$ 96.1, found 96.2.

HRMS (ESI): m/z calculated for $[IMe\cdot H]^+$ 97.0760, found 97.0760.

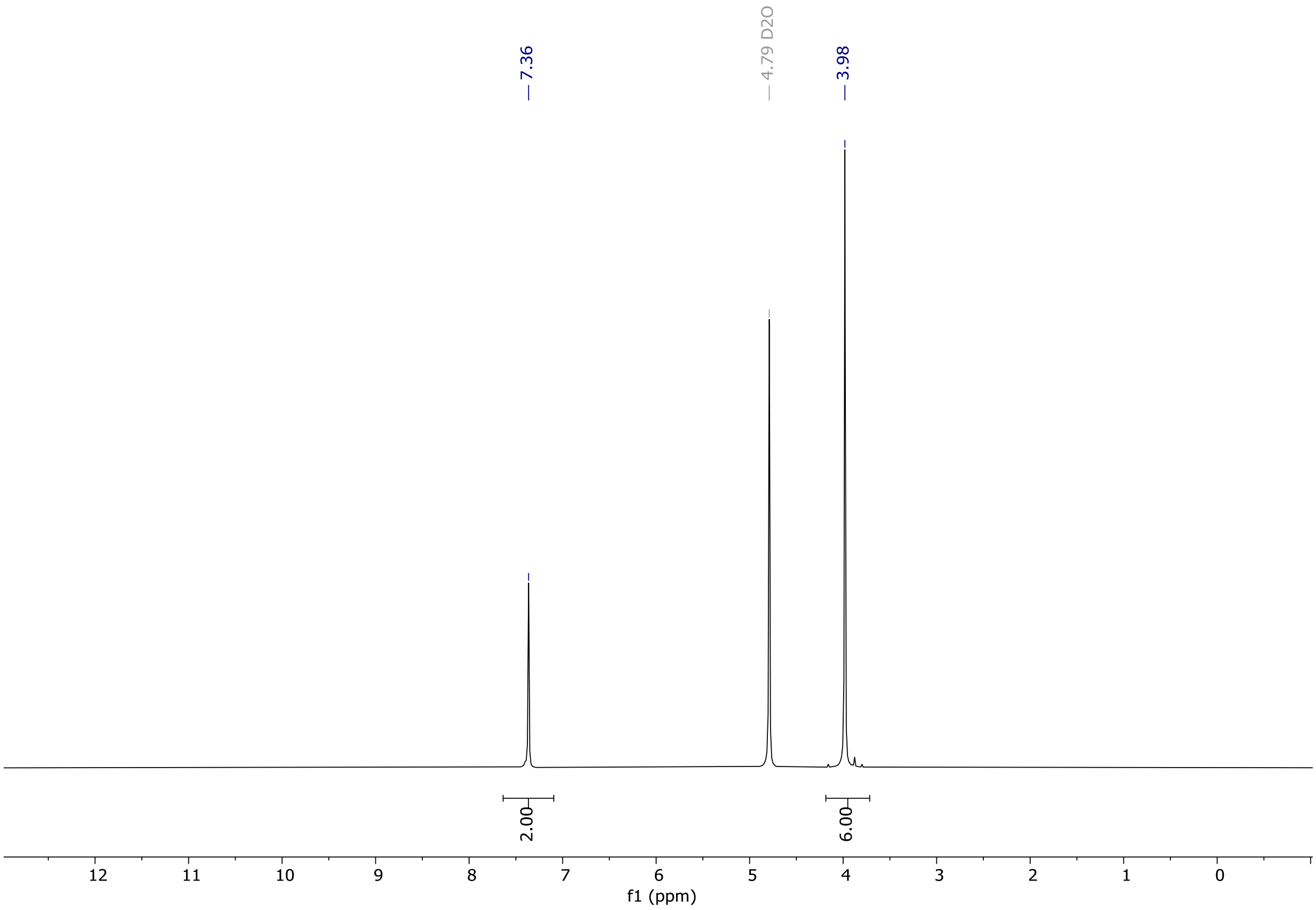

**Figure S9:** $^{1}H$ NMR (400 MHz, $D_2O$) spectrum of IMe·$CO_2$.

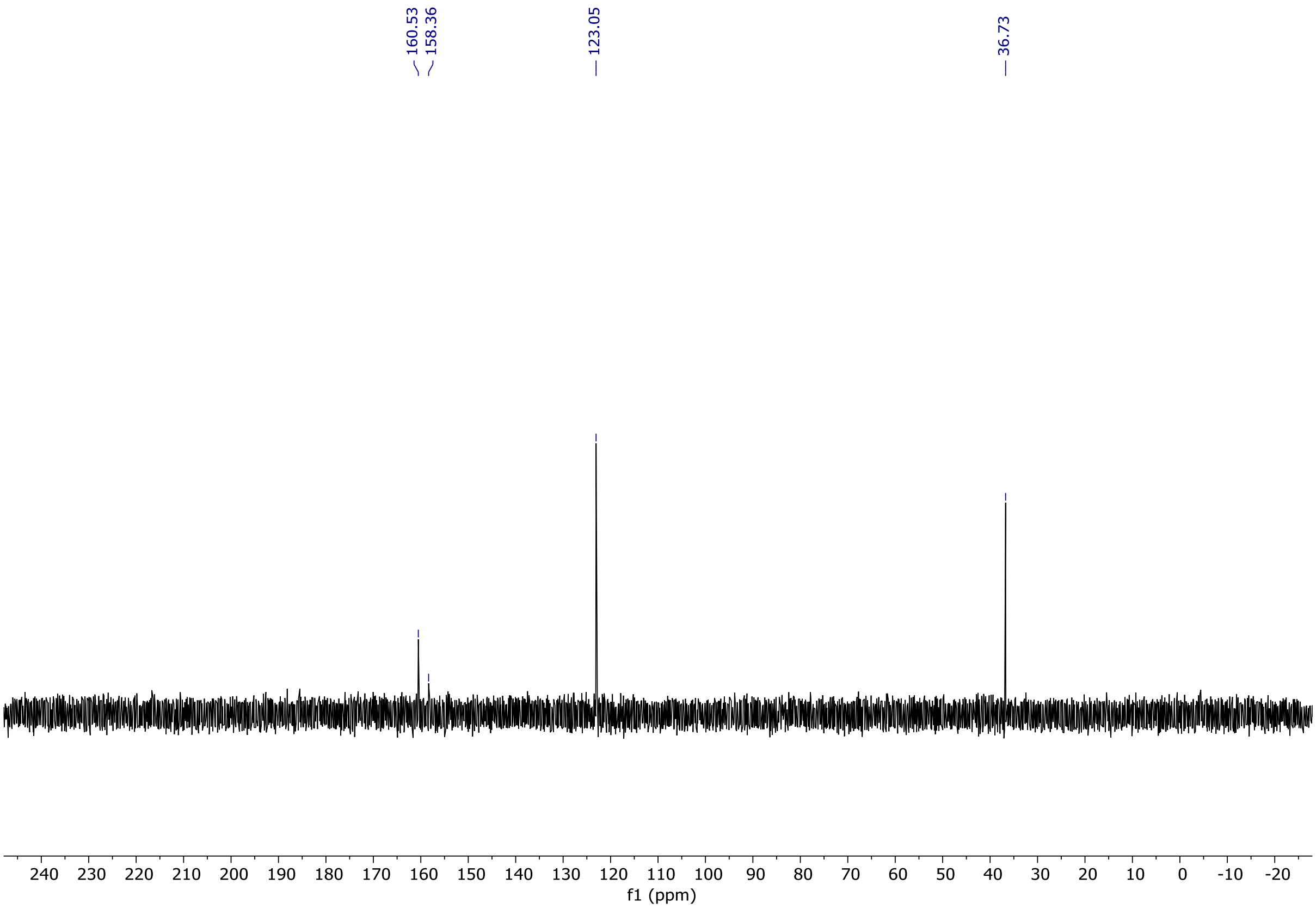

**Figure S10:** $^{13}C$ NMR (101 MHz, $D_2O$) spectrum of IMe·$CO_2$.

## Thermal Annealing and Physical Vapor Deposition

The annealing chamber and the deposition chamber were placed side by side in a glove-bag, which was filled with dry nitrogen (evaporated from liquid phase) and operated under constant nitrogen flow. Annealing was conducted at $10^{-5}$ mbar pressure at 190°C for approximately one hour. The annealed substrates were transferred from the annealing chamber into the deposition chamber without leaving the inert nitrogen atmosphere.

Physical vapor phase deposition (PVD) was carried out with a house-made deposition chamber (see Figure S11) which was kept inside the glovebag in nitrogen atmosphere during operation. In order to clean the chamber, its valve was connected to a vacuum pump (0.1-1.0 mbar) whilst the setup was heated to T > 150 °C for one hour. After cooling down to room temperature, the chamber was flooded with nitrogen, 2 mg of IMe·$CO_2$ was loaded in the crucible and two substrates were placed on the sample stage. One substrate was gold (Au) and the other Si/$SiO_2$ with exfoliated $MoS_2$ flakes of varying thickness. The chamber was flushed with nitrogen, evacuated (0.1 mbar) and the valve was closed. Now, the setup was heated to 120 °C for 30 minutes (p < 5 mbar). During cooling down, vacuum was restored upon reaching T < 80°C. At room temperature, the chamber was flushed with nitrogen to take out the samples.

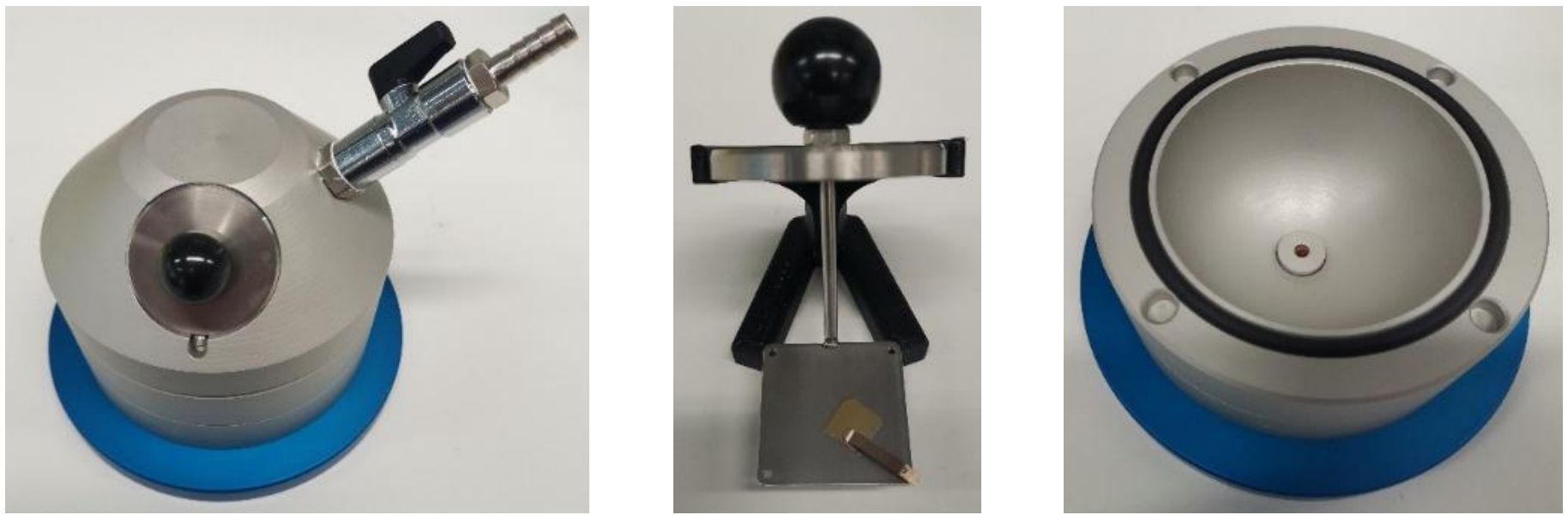

**Figure S11:** Physical vapor phase deposition chamber made from aluminum with Teflon sealings. The bottom part is compatible with standard laboratory heating plates. It contains a sample stage, a small crucible and a vacuum/nitrogen valve.

## XPS Analysis

The XPS measurement parameters and spectra are shown in Table S1 and Figure S12, respectively. All spectra were referenced relative to the Au $4f_{7/2}$ peak at a binding energy (BE) of 84 eV. An Al $K_\alpha$ X-ray source was used (12 kV filament, hν = 1486.7 eV), the spot size was 100 μm (72 W max), X-ray incidence and take-off angle amount to 60° and 0°, respectively, relative to surface normal. For charge compensation metallic clips on thin film connected to grounded sample stage and an argon flood gun were used.

**Table S1**: XPS measurement parameters.

| | Pass energy [eV] | Dwell time [ms] | Step size [eV] | No. of scans |
|---|---|---|---|---|
| Survey | 200 | 100 | 0.90 | 2 |
| Au4f | 50 | 100 | 0.05 | 5 |
| N1s | 100 | 100 | 0.05 | 10 |
| C1s | 50 | 50 | 0.05 | 5 |
| O1s | 50 | 50 | 0.05 | 5 |

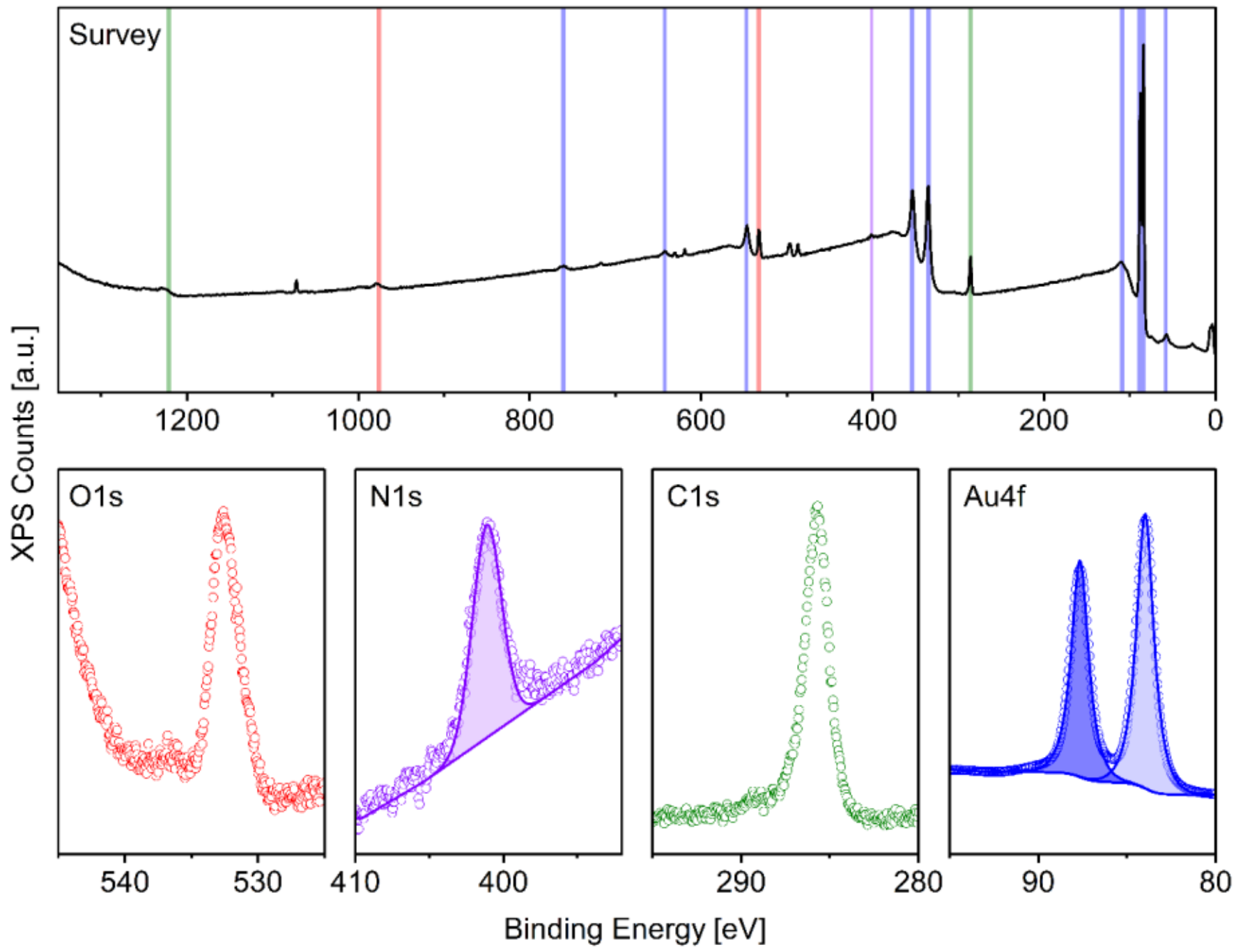

**Figure S12:** XPS spectra of IMe on Au. Peaks in the survey spectrum were marked with color according to the respective element as annotated in the atomic orbital spectra below: Oxygen red, nitrogen purple, carbon green and gold blue.

The presence of the N 1s signal is a clear indication of chemisorbed IMe on Au since it's N 1s BE of 401.1 eV agrees well with previous reports.[1] However, the successful binding of IMe to Au is solely an indication for the general working principle of the experimenal PVD procedure. It solely proofs that free carbenes have been present in the gas phase. At this state, no assumption can be made about IMe binding to $MoS_2$.

We also attempted XPS measurements of pristine and NHC-modified $MoS_2$ flakes. However, the small lateral dimensions of the flakes, combined with the intrinsically low sensitivity factor of the N 1s core level, resulted in an insufficient signal-to-noise ratio. In addition, the Mo 3p emission overlaps with the N 1s region, preventing reliable separation and quantification of the two contributions. To avoid overinterpretation, the corresponding XPS data and analysis are therefore not included.